\documentclass[11pt,a4paper]{article}
\usepackage[nosort]{cite}
\usepackage{amsmath,amsfonts,amssymb}
\usepackage[small,bf,hang]{caption}
\usepackage{slashed}
\usepackage{latexsym,epsfig}

\usepackage{hyperref}

\def\hybrid{
        \topmargin -20pt
        \oddsidemargin 0pt
        \headheight 0pt \headsep 0pt
        \textwidth 6.25in 
        \textheight 9.5in 
        \marginparwidth .875in
        \parskip 5pt plus 1pt}
\hybrid

\def\moth{\mathsurround=0pt}
\newdimen\zo \zo=0pt

\def\tick{\leaders\hrule height 0.5ex depth 0pt \hskip 0.5pt}
\def\upboxfill{$\moth \setbox\zo\hbox{\tick}%
  \hskip 3pt\hbox to 0pt{$\tick$\hss}\hrulefill \hbox to 7.5pt{$\tick$\hss}$}

\def\dtick{\leaders\hrule height .34pt depth 0.5ex \hskip 0.5pt}
\def\downboxfill{$\moth \setbox\zo\hbox{\dtick}%
  \hskip 2pt\hbox to 0pt{$\dtick$\hss}\hrulefill \hbox to 2pt{$\dtick$\hss}$}

\def\bec{\begin{center}}
\def\ec{\end{center}}

\def\cK{{\cal K}}

\def\cA{{\cal A}}

\def\cM{{\cal M}}
\def\cN{{\cal N}}
\def\cR{{\cal R}}

\def\cQ{{\cal Q}}

\def\cA{{\cal A}}

\def\be{\begin{equation}}
\def\ee{\end{equation}}
\def\bea{\begin{eqnarray}}
\def\eea{\end{eqnarray}}
\def\ba{\begin{array}}
\def\ea{\end{array}}

\newcommand{\CK}{\cK}
\newcommand{\CM}{\cM}
\newcommand{\CN}{\cN}
\newcommand{\CP}{{\cal P}}
\newcommand{\CQ}{\cQ}
\newcommand{\CR}{\cR}
\newcommand{\CA}{\cA}

\newcommand{\CFK}{\CK}
\newcommand{\CFM}{\CM}
\newcommand{\CFN}{\CN}
\newcommand{\CFP}{\CP}
\newcommand{\CFQ}{\CQ}
\newcommand{\CFR}{\CR}

\newcommand{\Bk}{k}
\newcommand{\Bm}{m}
\newcommand{\Bn}{n}

\newcommand{\EM}{M}
\newcommand{\EN}{N}

\newcommand{\adA}{\Sigma}

\newcommand{\talpha}{\alpha}
\newcommand{\tbeta}{\beta}

\newcommand{\Ua}{\underline{k}}
\newcommand{\Ub}{\underline{m}}
\newcommand{\Uc}{\underline{n}}

\newcommand{\falpha}{I}
\newcommand{\fbeta}{J}
\newcommand{\fgamma}{K}
\newcommand{\fdelta}{L}

\newcommand{\um}{u}
\newcommand{\un}{v}
\newcommand{\uk}{w}
\newcommand{\ul}{x}
\newcommand{\up}{y}
\newcommand{\uq}{z}

\newcommand{\La}{a}
\newcommand{\Lb}{b}
\newcommand{\Lc}{c}

\newcommand{\CX}{{\cal X}}

\begin{document}

\begin{center}

\vskip 5cm
\begin{flushright}
LMU--ASC 62/17
\end{flushright}

\vskip 1.5cm

{\Large \bf Ten-dimensional origin of \\[1ex]
Minkowski vacua in ${\cal N}=8$ supergravity}

\vskip 1cm

{\large \bf Emanuel Malek\,$^a$} {\large and} {\large \bf Henning Samtleben\,$^b$} \\

\vskip 25pt

{\em  
$^a$ Arnold Sommerfeld Center for Theoretical Physics, \\
Department f\"ur Physik, Ludwig-Maximilians-Universit\"at M\"unchen,\\ Theresienstra{\ss}e 37, 80333 M\"unchen, Germany.\\
{\tt E.Malek@lmu.de}\\[2ex]

$^b$ Univ Lyon, Ens de Lyon, Univ Claude Bernard, CNRS,\\
Laboratoire de Physique, F-69342 Lyon, France.}\\
{\tt Henning.Samtleben@ens-lyon.fr} \\

\end{center}

\vskip 1cm

\begin{abstract}

\noindent 
Maximal supergravity in four dimensions admits two inequivalent dyonic gaugings of the group SO(4)$\times$SO(2,2)$\ltimes T^{16}$. Both admit a Minkowski vacuum with residual SO(4)$\times$SO(2)$^2$ symmetry and identical spectrum. We explore these vacua and their deformations. Using exceptional field theory, we show that the four-dimensional theories arise as consistent truncations from IIA and IIB supergravity, respectively, around a Mink$_4\times S^3\times H^3$ geometry. The IIA/IIB truncations are efficiently related by an outer automorphism of SL(4) $\subset$ E$_{7(7)}$. As an application, we give an explicit uplift of the moduli of the vacua into a 4-parameter family of ten-dimensional solutions.

\end{abstract}

\vskip 1cm

\section{\mbox{Introduction: $D=4$ Minkowski vacua from ten dimensions}}

Maximal ${\cal N}=8$ gauged supergravity in four dimensions allows for a number
of Minkowski vacua with various gauge groups and 
different degrees of supersymmetry, many of which 
have only been revealed and studied in recent 
years~\cite{Cremmer:1979uq,Hull:2002cv,DallAgata:2011aa,DallAgata:2012tne,Catino:2013ppa}.
Their existence is often based on symplectic deformations
of maximal supergravity~\cite{DallAgata:2014ita,Inverso:2015viq}
whose higher-dimensional origin in turn remains largely mysterious.

In an a priori unrelated development, new efficient tools for the higher-dimensional
uplift of four-dimensional solutions and theories have emerged from the 
duality covariant reformulations of the higher-dimensional supergravity theories. 
In this framework, non-toroidal compactifications of supergravity are 
realized as generalized Scherk-Schwarz reductions on extended 
spacetimes~\cite{Scherk:1979zr,Aldazabal:2011nj,
Geissbuhler:2011mx,Berman:2012uy,Lee:2014mla,Hohm:2014qga,Inverso:2017lrz}.
In~\cite{Baguet:2015iou}, these techniques were used to prove a conjecture from \cite{Duff:1986ya}
that the NS-NS sector of ten-dimensional supergravity admits
a consistent truncation based on a group manifold $G$ 
to a half-maximal supergravity retaining non-abelian gauge 
bosons associated with the full isometry group $G\times G$\,.
The scalar fields of the lower-dimensional theory parametrize
the coset space ${\mathbb{R}}^+\times {\rm SO}(d,d)/({\rm SO}(d)\times {\rm SO}(d))$ 
with $d={\rm dim}\,{G}$ and couple via the scalar potential
\bea
V &=& \frac1{12}\,e^{2\,\varphi(x)}
X_{\CFM \CFN}{}^{\CFK} X_{\CFP \CFQ}{}^{\CFR} M^{\CFM \CFP}(x) \left( M^{\CFN \CFQ}(x) M_{\CFK\CFR}(x) + 
3\,\delta_{\CFK}^{\CFQ}\delta_{\CFR}^{\CFN}\, \right)
\;.
\label{Vpot}
\eea
Here, $M_{\CFM\CFN}(x)$ is the ${\rm SO}(d,d)$ valued matrix parametrizing the 
scalar target space, $\varphi(x)$ is the dilaton field, and the 
generalized structure constants $X_{\CFM\CFN}{}^{\CFK}$ encode 
the structure constants $f_{\Bk\Bm\Bn}$ of the group $G$, 
see~(\ref{XABC}) below.

It has further been observed in~\cite{Baguet:2015iou}
that for non-compact groups $G$
the potential (\ref{Vpot}) admits a Minkowski vacuum if the
number of compact and non-compact generators of ${G}$ 
are related by $n_{\rm cp}=2\,n_{\mbox{\scriptsize non-cp}}$\,.
An interesting example of such a group 
which we shall further study in this paper is
provided by 
\bea
{G}&=& {\rm SO}^*(4)~\equiv~ {\rm SO}(3)\times {\rm SO}(2,1)
\;,
\label{G4}
\eea
which gives rise to a four-dimensional ${\cal N}=4$ supergravity with gauge group
\bea
{\rm G}_{\rm gauge} &=& {G}\times{G}~=~
{\rm SO}(4)\times{\rm SO}(2,2)
\;,
\label{gauge-ss}
\eea
embedded into the isometry group of the
scalar target space ${\rm SO}(6,6)/({\rm SO}(6)\times {\rm SO}(6))$.

The associated Minkowski vacuum of the scalar potential (\ref{Vpot})
corresponds to a ten-dimensional solution of the type
\bea
{\rm Mink}_4 \times S^3 \times H^3
\label{solD10}
\;,
\eea
of a warped product of four-dimensional Minkowski space, a compact three-sphere,
and the non-compact hyperboloid $H^3$ with isometry group ${\rm SO}(2,2)$\,.
At the vacuum, the gauge group (\ref{G4}) is broken down to its compact part,
${\rm SO}(4)\times{\rm SO}(2)\times{\rm SO}(2)$, and supersymmetry is
completely broken. Yet the vacuum is classically stable at the quadratic level~\cite{DallAgata:2011aa}.

The aim of this letter is to further explore the Minkowski vacuum (\ref{solD10}) and its deformations. 
We construct its embedding into ${\cal N}=8$ supergravity, i.e.\ into  
consistent truncations of IIA and IIB supergravity
to inequivalent ${\cal N}=8$ gauged supergravities, both gauging the same non-semisimple group
\bea
{\rm SO}(4)\times{\rm SO}(2,2) \ltimes T^{16}
\;,
\label{gauge-full}
\eea
extending (\ref{gauge-ss}).
To show the inequivalence of the gaugings, we work out and compare their scalar potentials 
in a 10-scalar truncation. We construct the twist matrices that allow an explicit uplift of the
four-dimensional theories into IIA and IIB supergravity, respectively, via a generalized Scherk-Schwarz reduction. Exceptional field theory is particularly useful for this because it captures both IIA and IIB supergravity in one formalism \cite{Blair:2013gqa,Hohm:2013vpa,Hohm:2013uia}.
As a further application, we give an explicit
uplift of the moduli of this vacuum into a 
4-parameter family of ten-dimensional solutions.
These deform the background geometry (\ref{solD10}) such that
only a ${\rm U}(1)^4$ subgroup of its isometries is preserved.

The rest of the letter is organized as follows. In section 2 we describe the inequivalent
embeddings of the half-maximal supergravity with gauge group (\ref{gauge-ss}) into 
${\cal N}=8$ supergravity. In section 3 we construct the twist matrices that describe the
uplift into IIA and IIB supergravity via generalized Scherk-Schwarz reduction of
 exceptional field theory. We illustrate the inequivalence of the two resulting four-dimensional theories
 by comparing their potentials in a 10-scalar truncation in section~4. Finally, in section 5 we give
 an explicit uplift of the moduli of the Minkowski vacuum into a 4-parameter
 solution of $D=10$ supergravity.

\section{Embedding into maximal supergravity}

In this section we  discuss the embedding of the $D=4$, ${\cal N}=4$ gauged supergravity
with gauge group (\ref{gauge-ss}) obtained from compactification on (\ref{solD10})
into ${\cal N}=8$ gauged supergravities describing
consistent truncations of maximal IIA and IIB supergravity, respectively.
We first discuss this embedding on the level of the four-dimensional supergravities
in terms of the embedding tensor, enhancing the gauge group (\ref{gauge-ss}) to (\ref{gauge-full}).
The latter gaugings have been found and studied in \cite{DallAgata:2011aa,DallAgata:2012tne,Catino:2013ppa,DallAgata:2014ita}.
We then review the consistent truncation of $D=10$, ${\cal N}=1$ supergravity
around the solution (\ref{solD10}) by virtue of a generalized Scherk-Schwarz
reduction encoded in a properly chosen ${\rm SO}(6,6)$ twist matrix $U$. Upon embedding of
this twist matrix into E$_{7(7)}$ we arrive at consistent truncations of IIA and IIB
supergravity to the ${\cal N}=8$ gauged supergravities.

\subsection{Embedding ${\cal N}=4$ into ${\cal N}=8$ supergravity}
\label{subsec:embedding}

The scalar potential (\ref{Vpot}) appears in a gauging of $D=4$, ${\cal N}=4$ supergravity
\cite{Kaloper:1999yr,Schon:2006kz} whose generalized structure constants $X_{\CFM\CFN}{}^{\CFK}$ 
are given in terms of the structure constants $f_{\Bk\Bm\Bn}$ of the group ${G}={\rm SO}^*(4)$ as
\bea
X_{\CFM\CFN}{}^{\CFK}&:&\;
X_{\Bk\Bm\Bn} = f_{\Bk\Bm\Bn} \;,\;\;
X_{\Bk}{}^{\Bm\Bn} = f_{\Bk}{}^{\Bm\Bn}\;,\;\;
X^{\Bk}{}_{\Bm}{}^{\Bn} = f^{\Bk}{}_{\Bm}{}^{\Bn}\;,\;\;
X^{\Bm\Bn}{}_{\Bk} = f^{\Bm\Bn}{}_{\Bk} 
\;,\;
\label{XABC}
\eea
where ${\rm SO}(6,6)$ indices $\CFM, \CFN = 1, \ldots, 12$, are decomposed as $\{V^{\CFM}\}\rightarrow \{V^{\Bm}, V_{\Bm}\}$ and raised/lowered with the ${\rm SO}(6,6)$ invariant $\eta_{\CFM\CFN}$, and adjoint algebra indices $\Bm, \Bn = 1, \ldots, 6$, are raised and lowered with the Cartan-Killing form $\kappa_{\Bm\Bn}$\,. 

To describe the embedding of this half-maximal into maximal supergravity, 
we consider the decomposition of the symmetry group of ungauged ${\cal N}=8$ supergravity
\bea
{\rm E}_{7(7)} &\longrightarrow& {\rm SO}(6,6) \times {\rm SL}(2)
\;,
\label{ESS}
\eea
such that vector fields and the adjoint representation decompose as
\bea
{\bf 56} &\longrightarrow& {\bf (12,2)} + {\bf (32_s,1)}\;,\nonumber\\
{\bf 133} &\longrightarrow& {\bf (66,1)} +{\bf (1,3)} + {\bf (32_c,2)}
\;,
\label{56_133}
\eea
respectively, i.e. for $\EM = 1, \ldots, 56$ and $\adA = 1, \ldots, 133$,
\bea
A_\mu{}^{\EM} &\longrightarrow& \{ A_\mu{}^{\CM\,\talpha} , A_\mu{}^{\CA} \}
\;,\qquad
\CM=1, \ldots, 12\,,\;\;\talpha=\pm\;,
\nonumber\\
T_{\adA} &\longrightarrow& \{ t^{\CM\CN}, t^{(\talpha\tbeta)}, t^{\dot{\CA}\,\alpha} \}
\;,\qquad
{\CA}, \dot{\CA}=1, \ldots, 32 \;.
\label{AT}
\eea
The gauge couplings in the maximal theory are described by an embedding tensor
$\Theta_{\EM}{}^{\adA}$ in the ${\bf 912}$ of E$_{7(7)}$~\cite{deWit:2002vt}
\bea
D_\mu &=& \partial_\mu + A_\mu{}^{\EM}\,\Theta_{\EM}{}^{\adA}\,T_{\adA}
\;.
\eea

In our case, $\Theta_{\EM}{}^{\adA}$ is induced by the 
embedding tensor $X_{\CFM\CFN\CFK}$, (\ref{XABC}), of the half-maximal ${\cal N}=4$ theory, living in the $({\bf 220,2})$
of ${\rm SO}(6,6) \times {\rm SL}(2)$, see~\cite{Dibitetto:2011eu} for a detailed discussion
of such embeddings. 
In particular, (\ref{XABC}) satisfies the additional quadratic constraints
\cite{Aldazabal:2011yz,Dibitetto:2011eu}
\bea
X_{\CFM\CFN\CFK}\,X^{\CFM\CFN\CFK} &=& 0\;,
\label{con_extra}
\eea 
required for an embedding into a maximal theory.\footnote{This
is a general property of ${\cal N}=4$ gaugings that descend from
Scherk-Schwarz reductions respecting the section constraints~\cite{Geissbuhler:2011mx}.}
In the maximal theory, this induces gauge couplings
\bea
A_\mu{}^{\EM}\,\Theta_{\EM}{}^{\adA}\,T_{\adA}
&=&
A_\mu{}^{\CFK+} X_{\CFK\CFM\CFN}\, t^{\CFM\CFN} 
+ A_\mu{}^{\CA} X_{\CFK\CFM\CFN}\, \Gamma^{\CFK\CFM\CFN}_{\CA\dot{\CA}} \, t^{\dot{\CA}+}
\;.
\label{ATT}
\eea
The first term describes the gauging of the 
$\mathfrak{so}(4)\oplus{\mathfrak{so}}(2,2)$ generators within the 
algebra $\mathfrak{so}(6,6)=\left\langle t^{\CFM\CFN}\right\rangle$, i.e.\ 
reproduces the gauge group (\ref{gauge-ss}) of the ${\cal N}=4$ theory. The second term,
which carries the ${\rm SO}(6,6)$ gamma-matrices $\Gamma^{\CFK\CFM\CFN}_{\CA\dot{\CA}}$
describes the new generators that are gauged in the maximal theory.
In our case, it is straightforward to see, that these correspond to 16 commuting generators
of ${\rm E}_{7(7)}$ that transform as a bi-fundamental vector under the semi-simple
part (\ref{gauge-ss}) of the gauge group.
The full gauge group within the maximal theory then is given by
\bea
{\rm G}_{\rm gauge} &=& \left( {\rm SO}(4) \times {\rm SO}(2,2) \right) \ltimes T^{16}
\;.
\label{gauge-max}
\eea

A closer analysis of the gauge couplings (\ref{ATT}) shows that the gauging of the 16
nilpotent generators can be realized in two different ways depending on if the higher-dimensional
origin corresponds to the IIA or the IIB theory. While the embedding of ${\rm SO}(6,6)$ into ${\rm E}_{7(7)}$
according to (\ref{ESS})
is unique, the subgroup ${\rm GL}(6)\subset {\rm SO}(6,6)$ can be embedded in two inequivalent ways,
related by an exchange of the ${\rm SO}(6,6)$ spinor representations, and corresponding to a IIA or IIB origin.
Accordingly, there are two ways of embedding the ${\cal N}=4$ theory into an ${\cal N}=8$
gauging with gauge group (\ref{gauge-max}).
Specifically, the additional 32 vector fields in (\ref{AT}) transforming in the spinor representation of ${\rm SO}(6,6)$ 
decompose as
\bea
\mbox{IIA} &:& \left\{
A_\mu{}^{0}\,,\;
A_\mu{}^{mn}\,,\;
A_\mu{}_{\,mn}\,,\;
A_\mu{}_{\,0}\,
\right\}\;,\nonumber\\
\mbox{IIB} &:& \left\{
A_\mu{}^{m}\,,\;
A_\mu{}^{kmn}\,,\;
A_\mu{}_{m}\,
\right\}\;,
\label{GL6}
\eea
under ${\rm GL}(6)$, with $m,n = 1, \dots, 6$\,.
In terms of the structure constants (\ref{XABC}), the couplings (\ref{ATT}) of these fields organize
according to
\bea
\mbox{IIA} ~:~ A_\mu{}^{\CA} X_{\CFK\CFM\CFN}\, \Gamma^{\CFK\CFM\CFN}_{\CA\dot{\CA}} \, t^{\dot{\CA}+} &=&
A_\mu{}_{\,0}\,f_{kmn}\,t^{kmn} + 
A_\mu{}_{\,kl}\left(f_{mnp} \,\varepsilon^{klmnpq}\, t_q 
+ f^{kl}{}_m\,t^m\right)
\nonumber\\
&&{}
+A_\mu{}^{mn} f^{kl}{}_p\, \varepsilon_{klmnrs}\,t^{prs}
\nonumber\\
&\equiv&
A_\mu{}_{\,0}\,\CX^0 + 
A_\mu{}_{\,kl}\,\CX^{kl} 
+A_\mu{}^{mn} \,\CX_{mn}
\;,
\nonumber\\[1ex]
\mbox{IIB} ~:~ A_\mu{}^{\CA} X_{\CFK\CFM\CFN}\, \Gamma^{\CFK\CFM\CFN}_{\CA\dot{\CA}} \, t^{\dot{\CA}+} &=&
A_\mu{}^{kmn}\left( f_{kmn}\,t_0 + f^{pq}{}_k \,\varepsilon_{pqrsmn}\,t^{rs} \right)
\nonumber\\
&&{}
+ A_\mu{}_{q} f_{mnp} \,\varepsilon^{klmnpq}\, t_{kl} + A_\mu{}^m f^{kl}{}_m\,t_{kl}
\nonumber\\
&\equiv&
A_\mu{}^{kmn}\,\CX_{kmn} + A_\mu{}_{q} \,\CX^q + A_\mu{}^m \CX_m
\;,
\label{ATAB}
\eea
where the generators $t^{\dot{\CA}+}$ decompose as (\ref{GL6}) (with IIA and IIB interchanged).

Although both expression seem to formally involve more than 16 vector fields and generators,
both, the IIA and the IIB connection can be shown to contain precisely 16 independent vector fields.
For example, the generators $\CX^{kl}$ and $\CX_{mn}$, etc., contracting the vector fields,
are not independent, but constrained by
\bea
\CX_{mn}\,\CX^{mn} &=& 0\;,
\nonumber\\
\varepsilon^{klmnpq}\,\CX_{klm}\,\CX_{npq} &=& 0 ~=~ \CX^m\,\CX_m
\;,
\eea
as follows from the Jacobi identities of the structure constants $f_{\Bm\Bn}{}^{\Bk}$\,.
As a result, for both cases in (\ref{ATAB}), the resulting gauge algebra is
identical to (\ref{gauge-max}), yet the two gaugings are inequivalent as we shall
explicitly confirm below by comparing their scalar potentials.

In \cite{DallAgata:2011aa}, two gaugings of maximal supergravity with 
gauge group (\ref{gauge-max}) have been identified, constructed in the ${\rm SL}(8)$
frame and in the ${\rm SU}^*(8)$ frame of E$_{7(7)}$, respectively. We will establish the link in 
section~\ref{sec:twist}, with the former one describing the IIB embedding and the
latter one describing the IIA embedding of the ${\cal N}=4$ theory.

\subsection{Uplift of ${\cal N}=4$ supergravity}

We have described the embedding of the ${\cal N}=4$ theory with embedding tensor (\ref{XABC})
into maximal ${\cal N}=8$ supergravity. The half-maximal theory can be obtained as a consistent
truncation from ten-dimensional supergravity. This is most conveniently described by a 
Scherk-Schwarz reduction in a double field theory (DFT) reformulation~\cite{Siegel:1993th,Hull:2009mi,Hohm:2010jy} of ten-dimensional supergravity,
in terms of an ${\rm SO}(6,6)$ twist matrix $U$ given by \cite{Baguet:2015iou}
\bea
U_M{}^{\underline{K}} &=& 
\left \{   -\,\kappa^{\underline{KL}}\,{\cal K}_{\underline{L}\,m} + \eta^{\underline{KL}}\,{\cal K}_{\underline{L}}{}^n\,\tilde{C}_{nm}\,,\;  
\eta^{\underline{KL}}\,{\cal K}_{\underline{L}}{}^{m}  \right\}
\;,
\label{U}
\eea
in terms of the Killing vectors
\bea
{\cal K}_{\underline{K}}{}^m &\equiv& 
\{ L_{\underline{k}}{}^m+R_{\underline{k}}{}^m, 
L^{{\underline{k}}\,m}-R^{{\underline{k}}\,m} \}\;,
\label{Killing}
\eea
of left and right $G\times G$ isometries,
the Cartan-Killing form $\kappa^{\underline{KL}}$, the ${\rm SO}(6,6)$ invariant tensor $\eta^{\underline{KL}}$ and the two-form gauge potential $\tilde{C}_{mn}$ of the three-form flux on the group manifold $G$ defined by
\bea
3\,\partial_{[k}\tilde{C}_{mn]}&=& \tilde{H}_{kmn}
~\equiv~
-16\,f^{\Ua\Ub\Uc}\,L_{\Ua\,k} L_{\Ub\,m} L_{\Uc\,n}  ~=~ -16 \,f^{\Ua\Ub\Uc}\,R_{\Ua\,k} R_{\Ub\,m} R_{\Uc\,n} \;.
\eea
We refer to \cite{Baguet:2015iou} for details.

The construction applies to arbitrary groups $G$.
The fact that the relevant group (\ref{G4}) factorizes into two three-dimensional groups implies that the twist matrix $U$ only lives in the subgroup 
\bea
{\rm SL}(4)\times {\rm SL}(4)&\simeq&
{\rm SO}(3,3)\times {\rm SO}(3,3) ~\subset~ {\rm SO}(6,6)\;.
\eea
An equivalent presentation of the twist matrix (\ref{U}) can be given in terms of the explicit ${\rm SL}(4)$-valued twist matrices for $S^3$ and $H^3$ from \cite{Lee:2014mla,Hohm:2014qga}.

The twist matrix together with a generalized Scherk-Schwarz Ansatz allow us to derive the explicit uplift formulae of the four-dimensional ${\cal N}=4$ supergravity up to ten dimensions \cite{Baguet:2015iou}.
As an example, we can use these formulae to derive the ten-dimensional origin of the
four-dimensional Minkowski vacuum carried by the scalar potential (\ref{Vpot}) at the scalar origin.
This ten-dimensional background is conveniently described
by embedding the six-dimensional internal space into $\mathbb{R}^8$ via the coordinates
\bea
\{U^a, Y^a \}
\;,&& a=1, \dots, 4\;,
\nonumber\\
\mbox{with}\qquad 
U^aU^a &=&1~=~Y^a\eta_{ab}Y^b
\;,
\label{coordUY}
\eea
in terms of the ${\rm SO}(2,2)$ invariant metric $\eta_{ab}={\rm diag}\{-1,-1,1,1\}$\,.
The $D=10$ dilaton and metric then take the following form 
\bea
e^{\phi} &=&  \frac{1}{\sqrt{1+2\,y^2}}\;,\qquad y^2\equiv (Y^1)^2+(Y^2)^2\;,
\nonumber\\
ds^2&=&
e^{-\phi/2}\,\eta_{\mu\nu}\,dx^\mu\,dx^\nu+
2\,e^{-\phi/2}\,dU^a \,dU^a
+2\,e^{3\phi/2}\,dY^a\,dY^a
\;,
\label{D10metric}
\eea
with the four-dimensional Minkowski metric $\eta_{\mu\nu}$\,.
The geometry is a warped product (\ref{solD10}) with manifest 
isometry group ${\rm SO}(1,3)\times{\rm SO}(4)\times {\rm SO}(2)^2$.
The three-form flux takes the form
\bea
H_3 &=&24\,\left(
\omega_S
+
e^{4\phi}\,\omega_H
\right)
\;,
\label{D10H}
\eea
in terms of the canonical volume forms $\omega_S$ and $\omega_H$, of $S^3$ and $H^3$ 
given by \cite{Lee:2014mla}
\begin{equation}
 \omega_S = \frac1{3!} \,\varepsilon_{abcd}\, U^a \, dU^b\wedge dU^c\wedge dU^d \,, \qquad 
 \omega_H = \frac1{3!} \,\varepsilon_{abcd}\, Y^a \, dY^b\wedge dY^c\wedge dY^d
  \,,
\end{equation}
respectively.

\subsection{Embedding DFT into ExFT}

The construction can be extended to maximal supergravity by embedding the ten-dimensional
supergravity into E$_{7(7)}$ exceptional field theory (ExFT)~\cite{Hohm:2013uia}. 
This is the duality covariant formulation of maximal supergravity in which the fields 
are reorganised into E$_{7(7)}$ covariant objects living on an extended space of 56
coordinates $\{Y^{\EM}\}$ constrained by the strong section condition
\bea
(t_\adA)^{\EM\EN}\,\partial_{\EM}\otimes\partial_{\EN} &=& 0\;,
\label{section_E7}
\eea
with the E$_{7(7)}$ generators $(t_\adA)^{\EM\EN}$\,.
There are two inequivalent solutions to this condition which correspond
to selecting within the $\{Y^\EM\}$ six internal coordinates corresponding to
either IIA or IIB supergravity \cite{Hohm:2013uia}. Only the former set of coordinates may be
extended by a seventh coordinate without violating (\ref{section_E7}), corresponding
to $D=11$ supergravity.

Consistent truncations to maximal supergravities are described in exceptional field theory by generalized Scherk-Schwarz reductions in terms of E$_{7(7)}$ valued twist matrices $U$. Upon embedding the ${\rm SO}(6,6)$ twist matrix (\ref{U}) into E$_{7(7)}$, we thus obtain an embedding of the four-dimensional maximal supergravities discussed in section (\ref{subsec:embedding}) into ten dimensions. Although the embedding of ${\rm SO}(6,6)$ into E$_{7(7)}$ is unique, the two inequivalent ways of identifying coordinates (corresponding to the inequivalent embeddings of ${\rm GL}(6)$ into E$_{7(7)}$) result in two inequivalent ten-dimensional uplifts, into IIA and IIB supergravity, respectively. The corresponding coordinates are identified within the 56 internal coordinates of E$_{7(7)}$ ExFT as
\bea
{\rm IIA} ~:~
 {\bf 56} &\longrightarrow&
\boxed{{\mathbf6'}_{-4}}+ {\mathbf1}_{-3}+{\mathbf6}_{-2}+{\mathbf{15}}_{-1}+ \mathbf{15}'_{+1}+\mathbf6'_{+2}+1_{+3}+\mathbf6_{+4}\;,
\nonumber\\
{\rm IIB} ~:~{\bf 56} &\longrightarrow&
\boxed{({\mathbf6'},\mathbf1)_{-4}}+(\mathbf6,\mathbf2)_{-2}+(\mathbf{20},\mathbf1)_0 +(\mathbf6',\mathbf2)_{+2}+ (\mathbf6,\mathbf1)_{+4}
\;.
\label{coordinates}
\eea

While this construction provides a neat and compact proof for the existence of 
consistent uplifts of these four-dimensional supergravities, in practice the embedding of the twist matrix (\ref{U}) into E$_{7(7)}$ requires its evaluation in the spinor representations of the group ${\rm SO}(6,6)$ according to the decomposition of (\ref{56_133}) which is a somewhat cumbersome exercise. In the next section we thus give an alternative direct derivation of the full E$_{7(7)}$ twist matrices.

\section{The IIA/IIB twist matrices}
\label{sec:twist}

In~\cite{Inverso:2016eet}, twist matrices for the uplift of certain
dyonic ${\cal N}=8$ gaugings have been constructed after
decomposing the 56 coordinates in the ${\rm SL}(8)$ frame into
what we will refer to as `electric' and `magnetic' coordinates
\bea
\left\{Y^\EM\right\} &=& \left\{Y^{[AB]}, Y_{[AB]}\right\}\;,\qquad
A,B=1, \dots, 8
\;.
\label{sl8}
\eea
In these coordinates the section condition (\ref{section_E7}) takes the form
\bea
\partial_{AC} \otimes \partial^{BC}+\partial^{BC} \otimes \partial_{AC} &=&
\frac18\,\delta_A^B\left(\partial_{CD} \otimes \partial^{CD}
+\partial^{CD} \otimes \partial_{CD}\right)
\;,\nonumber\\
\partial_{[AB} \otimes \partial_{CD]} &=&
\frac1{24}\,
\varepsilon_{ABCDEFGH}\,\partial^{EF} \otimes \partial^{GH}
\;,
\label{section_SL8}
\eea
and twist matrices are constructed as products of matrices depending on
electric and on magnetic coordinates, respectively.

\subsection{IIB twist matrix}

Choosing physical coordinates as 
\bea
\left\{
y^i \equiv Y^{i8}\;,\quad \tilde{y}_{\La} \equiv Y_{\La7}
\right\}
\;,\qquad
i, j \in\{ 1, 2, 3\}\,,\; \La, \Lb \in\{ 4, 5, 6\}
\;,
\label{co_IIB}
\eea
among (\ref{sl8}), it is straightforward to verify that restricting the dependence of fields to these coordinates
solves the section condition (\ref{section_SL8}) and that (\ref{co_IIB}) cannot be extended by any of the other
50 internal coordinates without violating the section constraint.
ExFT evaluated on these coordinates thus describes IIB supergravity. Specifically, the 
${\rm GL}(1)_{\rm IIB}$, which provides the geometric grading of
coordinates (\ref{coordinates}) and fields, is generated by
\bea
{\rm GL}(1)_{\rm IIB} &=& \left\langle \frac34 \left(T_8{}^8 -T_7{}^7\right) +
\frac14 \left( T_1{}^1+T_2{}^2+T_3{}^3-T_4{}^4-T_5{}^5- T_6{}^6\right) \right\rangle
\;,
\eea
resulting in the charges 
\bea
\{Y^{i8}, Y_{\La7}\} : -4\,,\quad
\{Y^{\La8}, Y^{ij}, Y_{i7}, Y_{\La\Lb}\}  :  -2 \,,\quad
\{ Y^{i\La}, Y^{78}, Y_{78}, Y_{i\La} \} : 0 \;,\quad
\dots\;,
\eea
for the coordinates, in accordance with (\ref{coordinates}).
The twist matrices considered in \cite{Inverso:2016eet} are of the form
\bea
U(y^i,\tilde{y}_{\La})&\equiv& \mathring{U}(\tilde{y}_{\La})\,\hat{U}(y^i)
\;,
\eea
with the two commuting factors $\mathring{U}$ and $\hat{U}$ given by the
sphere/hyperboloid solutions from~\cite{Hohm:2014qga}.
They describe the embedding of maximal four-dimensional gaugings with
a dyonic embedding tensor given by
\bea
X_{AB,CD}{}^{EF} &=& \eta_{A[C} \delta_{D]B}{}^{EF} - \eta_{B[C} \delta_{D]A}{}^{EF} \;,
\nonumber\\
X^{AB}{}_{CD}{}^{EF} &=& -\tilde\eta^{A[E} \delta_{CD}{}^{F]B} + \tilde\eta^{B[E} \delta_{CD}{}^{F]A} \;,
\label{embeddingX}
\eea
with
\begin{equation}
\begin{split}
\eta_{AB} &= \mathrm{diag}(\overbrace{1,\ldots,1}^{p},\overbrace{-1,\ldots,-1}^{4-p},0,\ldots\ldots,0)\,,\\
\tilde\eta^{AB} &= \mathrm{diag}(0,\ldots\ldots,0,\underbrace{1,\ldots,1}_{q},\underbrace{-1,\ldots,-1}_{4-q})\,, \label{eq:IIBeta}
\end{split}
\end{equation}
as constructed in \cite{DallAgata:2011aa}.
For this paper, we are interested in the case $p=4$, $q=2$, corresponding to the gauge group (\ref{gauge-max}).
The gauge algebra thus is a subalgebra of $\mathfrak{sl}(8)=\langle T_A{}^B \rangle$ 
with the gauge connection given by
\bea
D_\mu &=& \partial_\mu -
\left(
A_\mu{}^{AB} \,\eta_{BC}
-A_{\mu\,CB} \,\tilde\eta^{BA}\right) T_A{}^C
\;,
\eea
corresponding to the IIB couplings of (\ref{ATAB}).

\subsection{IIA twist matrix}
We note that the above IIB Ansatz defines a natural embedding ${\rm SL}(4) \times {\rm SL}(4) \subset {\rm SL}(8)$, with each of the twist matrices $\mathring{U}$ and $\hat{U}$ an element of one of the two ${\rm SL}(4)$ factors. Using these two ${\rm SL}(4)$ subgroups we can slightly generalise the above Ansatz for the physical coordinates, by embedding them into
\begin{equation}
 \left\{ Y^{\falpha\fbeta}\,, \,\, Y_{\dot{\falpha}\dot{\fbeta}} \right\} \,, \qquad \falpha, \fbeta = \left\{ 1, 2, 3, 8 \right\} \,, \quad \dot{\falpha}, \dot{\fbeta} = \left\{ 4, 5, 6, 7 \right\} \,,
\end{equation}
with $\hat{U}(Y^{\falpha\fbeta})$ and $\mathring{U}(Y_{\dot{\falpha}\dot{\fbeta}})$. The section condition \eqref{section_SL8} then becomes
\begin{equation}
 \varepsilon^{\falpha\fbeta\fgamma\fdelta} \partial_{\falpha\fbeta} \otimes \partial_{\fgamma\fdelta} = \varepsilon_{\dot{\falpha}\dot{\fbeta}\dot{\fgamma}\dot{\fdelta}} \partial^{\dot{\falpha}\dot{\fbeta}} \otimes \partial^{\dot{\fgamma}\dot{\fdelta}} \,.
\end{equation}
Here, we further restrict to the case
\begin{equation}
 \varepsilon^{\falpha\fbeta\fgamma\fdelta} \partial_{\falpha\fbeta} \otimes \partial_{\fgamma\fdelta} = \varepsilon_{\dot{\falpha}\dot{\fbeta}\dot{\fgamma}\dot{\fdelta}} \partial^{\dot{\falpha}\dot{\fbeta}} \otimes \partial^{\dot{\fgamma}\dot{\fdelta}} = 0 \,.
\end{equation}

We can now follow \cite{Malek:2015hma} and apply the outer automorphism of the ${\rm SL}(4)$ factor defined by $\dot{\falpha}, \dot{\fbeta} = \left\{ 4, 5, 6, 7 \right\}$. This takes
\begin{equation}
  \partial_{\dot{\falpha}\dot{\fbeta}} \longrightarrow \frac12 \varepsilon^{\dot{\falpha}\dot{\fbeta}\dot{\fgamma}\dot{\fdelta}} \partial_{\dot{\fgamma}\dot{\fdelta}} \,, \qquad
  \mathring{U} \longrightarrow \mathring{U}^{-T} \,,
\end{equation}
and one can easily show that it satisfies the conditions in \cite{Inverso:2016eet}, ensuring a consistent truncation. The new full set of physical coordinates is now given by
\bea
\left\{
y^i \equiv Y^{i8}\;,\quad \tilde{y}^{\La} \equiv \frac12 \varepsilon^{\La\Lb\Lc}\,Y_{\Lb\Lc}
\right\}
\;,\qquad
i, j \in\{ 1, 2, 3\}\,,\; \La, \Lb \in\{ 4, 5, 6\}
\;,
\label{co_IIA}
\eea
with twist matrix
\bea
U(y^i,\tilde{y}^{\La})&\equiv& \mathring{U}(\tilde{y}^{\La})^{-T}\,\hat{U}(y^i)
\;.
\eea

It is straightforward to verify that the new set of coordinates can be extended by a seventh coordinate $Y^{78}$, while still satisfying the section constraints (\ref{sl8}). The resulting theory is thus type IIA supergravity (with possible $D=11$ embedding). The ${\rm GL}(1)_{\rm IIA}$, which provides the geometric grading of coordinates (\ref{coordinates}) and fields, is generated by
\bea
{\rm GL}(1)_{\rm IIA} &=& \left\langle 
\frac13 \left( T_1{}^1+T_2{}^2+T_3{}^3\right) 
-\frac23  \left(T_4{}^4 + T_5{}^5+ T_6{}^6\right) + T_8{}^8
\right\rangle
\;,
\eea
giving charges ($i, j = 1, \dots, 3$, $\La, \Lb = 4, \dots, 6$)
\bea
\{Y^{i8}, Y_{\La\Lb}\} : -4\,,\quad
\{Y^{78}\} : -3
\,,\quad
\{Y^{ij}, Y_{\La7}\} : -2
\,,\quad
\{Y^{i7}, Y^{\La8}, Y_{i\La}\} : -1\;,\quad
\dots\;,
\eea
for the coordinates, in accordance with (\ref{coordinates}).

The new embedding tensor corresponding to this IIA reductions is given by
\begin{equation}
 \begin{split}
 X_{AB,CD}{}^{EF} &= \eta_{A[C} \delta_{D]B}{}^{EF} - \eta_{B[C} \delta_{D]A}{}^{EF} \;, \\
 X^{AB}{}_{CD}{}^{EF} &= -\Sigma^{AB[E}_{[C} \delta_{D]}{}^{F]} \;,
 \end{split} \label{embeddingXIIA}
\end{equation}
where
\begin{equation}
 \Sigma^{ABC}_D = \omega^{ABCE} \tilde{\eta}_{DE} \,,
\end{equation}
and the only non-vanishing components of $\omega^{ABCD}$ are
\begin{equation}
 \omega^{\dot{\falpha}\dot{\fbeta}\dot{\fgamma}\dot{\fdelta}} = \varepsilon^{\dot{\falpha}\dot{\fbeta}\dot{\fgamma}\dot{\fdelta}} \,,
\end{equation}
and $\tilde{\eta}_{AB} = \tilde{\eta}^{AB}$ of \eqref{eq:IIBeta} with $p = 4$ and $q = 2$. The quadratic constraint for this type of embedding tensor are given by
\begin{equation}
 \begin{split}
  \Sigma_{(A}{}^{CDE} \eta_{B)E} &= 0 \,, \\
  \Sigma_A{}^{BC[D} \Sigma_C{}^{EFG]} &= 0 \,,
 \end{split}
\end{equation}
which are satisfied in the given case.

At this stage it is natural to ask whether the 4-dimensional gauged SUGRAs we obtained from IIA and IIB are different. In 7 dimensions, the IIA / IIB truncations related by an outer automorphism of ${\rm SL}(4)$ are clearly inequivalent because the resulting embedding tensor belongs to different irreducible representations under the global symmetry group ${\rm SL}(5)$ \cite{Malek:2015hma}. Here, this is much harder to assess because in both cases the embedding tensor belong to the $\mathbf{912}$ representation under ${\rm E}_{7(7)}$. Under ${\rm SL}(8) \subset {\rm E}_{7(7)}$, a difference emerges: the IIA embedding tensor corresponds to gaugings in the $\mathbf{36}$ and $\overline{\mathbf{420}}$ of ${\rm SL}(8)$ while the IIB one to gaugings in the $\mathbf{36}$ and $\overline{\mathbf{36}}$ of ${\rm SL}(8)$. However, the two embedding tensors also couple to different sets of vector fields so that this direct comparison is meaningless. Nonetheless, the IIA embedding tensor takes the same form as in \eqref{embeddingX} in the ${\rm SU}^*(8)$ frame, with gaugings in the $\mathbf{36}$ and $\overline{\mathbf{36}}$. This suggests that the IIA and IIB reductions yield different gauged SUGRAs, as one would have expected and as we will explicitly confirm in the next section.


\section{Gaugings and potentials}


So far we have shown that
IIA and IIB supergravity compactified around (\ref{D10metric})--(\ref{D10H}) give rise to maximal $D=4$ supergravities which share the same gauge group (\ref{gauge-max}) but embedded in inequivalent ways within E$_{7(7)}$\,.
Around this background the two theories exhibit the same spectrum as can be confirmed by expanding the resulting scalar potentials to quadratic order.

In order to confirm explicitly that the two gaugings represent 
inequivalent four-dimensional theories, 
we will compute and compare a truncation of their full
scalar potentials.
To this end, we consider their respective truncations to singlets under the compact subgroup
\bea
G_{0}~\equiv~{\rm SO}(3)_{\rm D} \times {\rm SO}(2)_{\rm D} &\subset&
\left({\rm SO}(3)\times{\rm SO}(3)\right)\times\left({\rm SO}(2)\times{\rm SO}(2)\right)
\nonumber\\
&\subset&
{\rm SO}(4)\times {\rm SO}(2,2)
\;,
\eea
of the gauge group. Within ${\rm E}_{7(7)}$ this group commutes with 
a ${\rm GL}(4) \times {\rm SO}(2)$,
i.e.\ the scalar coset ${\rm E}_{7(7)}/{\rm SU}(8)$ contains 10 singlets under $G_0$
with the resulting kinetic term given by
\bea
{\cal L}_{\rm scal} &=& \frac14\,
D_\mu M_{\um\un} D^\mu M^{\um\un}
+3\,D_\mu \lambda \,D^\mu \lambda
\;,
\eea
in terms of a symmetric ${\rm SL}(4)$ matrix $M_{\um\un}$, $\um, \un = 1, \ldots, 4$ and a scalar $\lambda$\,.
Under reduction to the common NS-NS sector, the ${\rm GL}(4)$ further breaks down 
to ${\rm GL}(2)\times {\rm GL}(2)$,
in particular the ${\rm SL}(4)$ matrix $M_{\um\un}$ breaks down to an
${\rm SL}(2)\times {\rm SL}(2) \times {\rm GL}(1)$ matrix of block-diagonal form
\bea
M_{\um\un} &=& 
\begin{pmatrix}
*&0\\
0&*
\end{pmatrix}
\;.
\eea

For a general gauging, the embedding tensor $\Theta_{\EM}{}^{\adA}$ in the ${\bf 912}$ representation
of E$_{7(7)}$ contains 64 singlets under $G_0$ which organise into ${\rm SL}(4)$ tensors according to
\bea
+3 &:& X_{[\um\un]}{}^\uk, Y^{[\um\un\uk]}, Z^{[\um\un\uk]}
\;,\nonumber\\
+1 &:&  f_{[\um\un\uk]} = \varepsilon_{\um\un\uk\ul}\,\tilde{f}^\ul
\;,\nonumber\\
-1 &:&   f^{[\um\un\uk]}= \varepsilon^{\um\un\uk\ul}\,\tilde{f}_\ul
\;,\nonumber\\
-3 &:& X^{[\um\un]}{}_\uk, Y_{[\um\un\uk]}, Z_{[\um\un\uk]}
\;,
\label{XXZZ}
\eea
with the grading referring to ${\rm GL}(1)$\,. The associated scalar potential
is computed by applying the truncation to $G_0$ singlets to the general ${\cal N}=8$
potential from \cite{deWit:2007mt}, resulting in
\bea
V &=& 
\frac1{16}\,e^{-3\lambda}\left(
X_{\um\un}{}^\uk X_{\up\uq}{}^\ul \,M_{\uk\ul} M^{\um\up} M^{\un\uq} + 2\,X_{\um\uk}{}^\ul X_{\un\ul}{}^\uk\, M^{\um\un} 
\right)
\nonumber\\
&&{}
-\frac38\,e^{-\lambda}\left(\tilde{f}^\um \tilde{f}^\un+f^{\um\uk\ul} X_{\uk\ul}{}^\un\right) M_{\um\un}
\nonumber\\
&&{}
+\frac14 \left(X_{\um\un}{}^\up Z_{\up\uk\ul}+X^{\um\un}{}_\up Z^{\up\uk\ul}\right)\,M^{\um\uk}M^{\un\ul}
\nonumber\\
&&
{}-\frac38\, e^{\lambda}\left(\tilde{f}_\um\tilde{f}_\un+f_{\uk\ul\um} X^{\uk\ul}{}_{\un}\right) M^{\um\un}
\nonumber\\
&&{}
+
\frac1{16}\,e^{3\lambda}\left(
X^{\um\un}{}_\uk X^{\up\uq}{}_\ul \,M^{\uk\ul} M_{\um\up} M_{\un\uq} + 2\,X^{\um\uk}{}_\ul X^{\un\ul}{}_\uk \,M_{\um\un} 
\right)
\;.
\label{genV}
\eea

For the IIA and IIB embedding tensors given in the last section, truncation
to $G_0$ singlets yields
\bea
\mbox{IIA:\;\;}
+3 &:& X_{\um\un}{}^\uk ~:~\{X_{23}{}^4=X_{42}{}^3=X_{34}{}^1=1 \}
\;,\nonumber\\
-1 &:&   f^{\um\un\uk}~:~\{f^{134}=1\}
\;,\nonumber\\
-3 &:& Z_{\um\un\uk}~:~ \{Z_{234}=1\}
\;;\label{IA}\\[1.5ex]
\mbox{IIB:\;\;}
+3 &:& X_{\um\un}{}^\uk~:~\{X_{23}{}^4=X_{42}{}^3=1 \}
\;,\nonumber\\
+1 &:&  f_{\um\un\uk}~:~\{f_{234}=1\}
\;,\nonumber\\
-3 &:& X^{\um\un}{}_\uk~:~\{X^{34}{}_2=1\}\;,\;\; Z_{mnk}~:~\{Z_{234}=1\}
\;,
\label{IB}
\eea
when written in the basis (\ref{XXZZ}). In particular, in this truncation, only an
${\rm SO}(2)\ltimes T^2$ subgroup of the gauge group (\ref{gauge-max}) survives
in both cases.
The inequivalence of the two resulting gaugings now becomes manifest from the 
different forms the general scalar potential (\ref{genV}) takes for (\ref{IA}) and (\ref{IB}),
respectively:
\bea
V_{\rm IIA} &=& 
\frac1{16}\,e^{-3\lambda}\left(
X_{\um\un}{}^k X_{\up\uq}{}^\ul \,M_{\uk\ul} M^{\um\up} M^{\un\uq} -4\,M^{22} 
\right)
-\frac34\,e^{-\lambda}\, M_{11}
\nonumber\\
&&{}
+\frac14\,X_{\um\un}{}^\up Z_{\up\uk\ul}\,M^{\um\uk}M^{\un\ul}
-\frac38\, e^{\lambda}\, M^{22}
\;,
\nonumber\\
V_{\rm IIB} &=& 
\frac1{16}\,e^{-3\lambda}\left(
X_{\um\un}{}^\uk X_{\up\uq}{}^\ul \,M_{\uk\ul} M^{\um\up} M^{\un\uq} -4\,M^{22} 
\right)
-\frac38\,e^{-\lambda}\, M_{11}
\nonumber\\
&&{}
+\frac14 \, X_{\um\un}{}^\up Z_{\up\uk\ul}\,M^{\um\uk}M^{\un\ul}-\frac34\, e^{\lambda}\, M^{22}
+
\frac1{8}\,e^{3\lambda}\,
M^{22} \left(M_{33} M_{44}-M_{34}M_{43}\right) 
\;.
\qquad
\eea
In particular, in the IIA potential  the $e^{3\lambda}$ term vanishes identically,
showing the inequivalent asymptotic behaviour of the two potentials.

\section{Uplift of the moduli}

Around the Minkowski vacuum, the four-dimensional theories have a six-dimensional moduli space~\cite{DallAgata:2012tne}
which can be identified within the NS-NS sector. Apart from the trivial $\frac{{\rm SL}(2)}{{\rm SO}(2)}$ factor from the ten-dimensional dilaton and 
Kalb-Ramond field, the remaining four moduli $\{\varphi_i, \chi_i\}$, $i = 1, 2$, form an 
$\left(\frac{{\rm SL}(2)}{{\rm SO}(2)}\right)^2 \subset {\rm SO}(6,6)$ embedded according to the 
scalar moduli matrix $M_{AB}$
\begin{equation}
M_{AB} = {\small
\left( \begin{array}{cccccccccccc}
 1 & 0 & 0 & 0 & 0 & 0 & 0 & 0 & 0 & 0 & 0 & 0 \\
 0 & 1 & 0 & 0 & 0 & 0 & 0 & 0 & 0 & 0 & 0 & 0 \\
 0 & 0 & s_1 s_2 & 0 & 0 & - s_1 \chi_2 & 0 & 0 & \chi_1 \chi_2 & 0 & 0 & s_2 \chi_1 \\
 0 & 0 & 0 & 1 & 0 & 0 & 0 & 0 & 0 & 0 & 0 & 0 \\
 0 & 0 & 0 & 0 & 1 & 0 & 0 & 0 & 0 & 0 & 0 & 0 \\
 0 & 0 & - s_1 \chi_2 & 0 & 0 & s_1 e^{-\varphi_2} & 0 & 0 & - e^{-\varphi_2} \chi_1 & 0 & 0 & - \chi_1 \chi_2 \\
 0 & 0 & 0 & 0 & 0 & 0 & 1 & 0 & 0 & 0 & 0 & 0 \\
 0 & 0 & 0 & 0 & 0 & 0 & 0 & 1 & 0 & 0 & 0 & 0 \\
 0 & 0 & \chi_1 \chi_2 & 0 & 0 & - e^{-\varphi_2} \chi_1 & 0 & 0 & e^{-\varphi_1-\varphi_2} & 0 & 0 & e^{-\varphi_1} \chi_2 \\
 0 & 0 & 0 & 0 & 0 & 0 & 0 & 0 & 0 & 1 & 0 & 0 \\
 0 & 0 & 0 & 0 & 0 & 0 & 0 & 0 & 0 & 0 & 1 & 0 \\
 0 & 0 & s_2 \chi_1 & 0 & 0 & - \chi_1 \chi_2 & 0 & 0 & e^{-\varphi_1} \chi_2 & 0 & 0 & s_2 e^{-\varphi_1}
\end{array} \right)} \,,
\end{equation}
with $s_{i} = e^{\varphi_{i}} \left(1+\chi_{i}^2\right)$.
Their kinetic term in the four-dimensional theory is given by
\bea
{\cal L}_{\rm kin}&=&
\frac12\left(
\partial_\mu \varphi_1\,\partial^\mu \varphi_1
+e^{-2\varphi_1}\,\partial_\mu \tilde\chi_1\,\partial^\mu \tilde\chi_1
+\partial_\mu \varphi_2\,\partial^\mu \varphi_2
+e^{-2\varphi_2}\,\partial_\mu \tilde\chi_2\,\partial^\mu \tilde\chi_2
\right)
\;,
\eea
with $\tilde\chi_i=e^{\varphi_i} \chi_i$\,. Using the formulae from \cite{Baguet:2015iou}
these moduli can be uplifted to 
$D=10$ dimensions and we will work out the explicit uplift here.

The ten-dimensional background corresponding to these massless deformations preserves a set of $U(1)^4$ isometries and is therefore most conveniently described in terms of the following functions on the six-dimensional internal space
\bea
&&\{u^\alpha, v^\alpha, y^\alpha, z^\alpha \}
\;, \quad\alpha=1,2\;,
\nonumber\\
&&\mbox{with} \quad
u^\alpha u^\alpha+v^\alpha v^\alpha ~=~1~=~
z^\alpha z^\alpha - y^\alpha y^\alpha
\;, \label{eq:ManEmbedding}
\eea
so that the $U(1)^4$ isometries are realised as rotations on the $\{u^\alpha, v^\alpha, y^\alpha, z^\alpha \}$, respectively. These functions are in fact the usual coordinates on $\mathbb{R}^8$ in which the six-dimensional manifold is embedded via \eqref{eq:ManEmbedding}. As a result these functions are \emph{globally} well-defined on the internal space and allow us to give global expressions for the metric and form fields on the internal space, rather than local coordinate expressions.

To this end we introduce the $U(1)^4$ invariant one-forms
\bea
\sigma_0 &\equiv& u^\alpha du^\alpha\;,\qquad
\sigma_1 ~\equiv~ \varepsilon_{\alpha\beta}\,u^\alpha du^\beta\;,\qquad
\sigma_2 ~\equiv~ \varepsilon_{\alpha\beta}\,v^\alpha dv^\beta\;,\nonumber\\
\tau_0 &\equiv& y^\alpha dy^\alpha\;,\qquad
\tau_1 ~\equiv~ \varepsilon_{\alpha\beta}\,y^\alpha dy^\beta\;,\qquad
\tau_2 ~\equiv~ \varepsilon_{\alpha\beta}\,z^\alpha dz^\beta\;,
\eea
and functions
$u^2\equiv u^\alpha u^\alpha$\,,
$y^2 \equiv y^\alpha y^\alpha$\,.
In terms of these forms, the volume forms, $\omega_{S/H}$, of the undeformed $S^3 / H^3$ are given by
\begin{equation}
 \omega_S = \frac12 \left( \sigma_1 \wedge d\sigma_2 + \sigma_2 \wedge d\sigma_1 \right) \,, \qquad \omega_H = \frac12 \left( \tau_1 \wedge d\tau_2 + \tau_2 \wedge d\tau_1 \right) \,,
\end{equation}
as can, for example, be seen from \cite{Lee:2014mla}.
We will moreover define the moduli-dependent functions
\bea
f_1(u) &\equiv&
e^{-\varphi_1}\left(1-u^2\right)+e^{\varphi_2}\,u^2\left(1+\chi_2^2\right)
\;,\nonumber\\
f_2(u) &\equiv&
e^{-\varphi_2}\left(1-u^2\right)+e^{\varphi_1}\,u^2\left(1+\chi_1^2\right)
\;,\nonumber\\
g_1(y) &\equiv&
e^{-\varphi_2}\,y^2+e^{-\varphi_1}\,\left(1+y^2\right)
\;,\nonumber\\
g_2(y) &\equiv&
e^{\varphi_1}\,y^2\left(1+\chi_1^2\right)+e^{\varphi_2}\,\left(1+y^2\right)
\left(1+\chi_2^2\right)
\;,
\eea
and note that for finite values of the moduli, these functions are given by a sum of two positive terms. Furthermore, those two terms do not both vanish at the same locations, and thus the functions $f_{i}$, $g_i$ are positive-definite for finite values of the moduli.

The $D=10$ metric yields a deformation of (\ref{D10metric})
\bea
ds^2&=&
\Delta^{-1}\,\eta_{\mu\nu}\,dx^\mu\,dx^\nu+
2\,e^{4\,\varphi_0}\,\Delta^3\,d\hat{s}^2_6
\;,
	\eea
with warp factor
\bea
\Delta^{-4} &=& e^{6\,\varphi_0}\left(
y^2\,f_2(u)+\left(1+y^2\right) f_1(u) \right)
\nonumber\\
&=& e^{6\,\varphi_0}\left(
 \left(1-u^2\right)g_1(y)
 +u^2\,g_2(y)
\right)
\;,
\eea
and an internal six-dimensional metric $d\hat{s}^2_6$
\bea
d\hat{s}^2_6 &=& 
g_1(y)\,
du^\alpha\,du^\alpha
+
g_2(y)\,
dv^\alpha\,dv^\alpha
+f_1(u)\,
dy^\alpha\,dy^\alpha
+
f_2(u)\,
dz^\alpha\,dz^\alpha
\nonumber\\
&&{}
+ 2\,\chi_2\left(\sigma_1\, \tau_1
+\sigma_2\, \tau_2\right)
 + 2\,\chi_1\left(
 \sigma_1\,\tau_2
+\sigma_2\,\tau_1\right)
\;.
\eea
The $D=10$ dilaton is given by
\bea
e^{\phi/2} &=& \Delta\,e^{2\,\varphi_0}
\;,
\eea
and the Kalb-Ramond form
\begin{equation}
 B = \tilde{B} + \imath_v \left( \omega_S + \omega_H \right) + 2 \Delta^4\, e^{6\,\varphi_0} \left[ \chi_1 \left( \sigma_1 \wedge \tau_1 + \sigma_2 \wedge \tau_2 \right) + \chi_2 \left( \sigma_1 \wedge \tau_2 + \sigma_2 \wedge \tau_1 \right) \right]  \,,
\end{equation}
where
\begin{equation}
  \tilde{H}_3 = d\tilde{B} = 4 \left( \omega_S + \omega_H \right) \,, \qquad
  v^i = 8 \Delta^{-1} \tilde{g}^{ij} \partial_j \Delta \,,
\end{equation}
and $\tilde{g}_{ij}$ is an auxiliary pseudo-Riemannian metric with line element
\begin{equation}
 d\tilde{s}_6^2 = du^\alpha du^\alpha + dv^\alpha dv^\alpha - dy^\alpha dy^\alpha + dz^\alpha dz^\alpha \,.
\end{equation}
The Kalb-Ramond three-form flux takes the form
\bea
H_3 &=&12\,e^{12\,\varphi_0}\,\Delta^8\,\hat{H}_3
\;,
\nonumber\\[1ex]
\mbox{with}\quad
\hat{H}_3&=&
2 g_1(y)\,g_2(y)\,
\omega_S
+
2 f_1(u)\,f_2(u)\,
\omega_H
\nonumber\\
&&{}
+g_1(y)\, 
d\sigma_1 \wedge 
\left(\chi_2\,\tau_2+\chi_1\,\tau_1\right)
+g_2(y)\,
d\sigma_2 \wedge 
\left(\chi_2\,\tau_1+\chi_2\,\tau_1\right)
\nonumber\\
&&{}
-
f_1(u) 
d\tau_1 \wedge 
\left(\chi_2\,\sigma_2+\chi_1\,\sigma_1\right)
+
f_2(u)\, 
d\tau_2 \wedge 
\left(\chi_2\,\sigma_1+\chi_1\,\sigma_2\right)
\nonumber\\
&&{}
+2\left(\chi_1^2-\chi_2^2\right)
\left(
\sigma_1\wedge\sigma_2\wedge \tau_0
-\tau_1\wedge\tau_2\wedge \sigma_0
\right)
\;,
\eea
describing a four-parameter deformation of (\ref{D10H}).

We have thus completed the uplift of the four moduli to the parameters of a
solution of $D=10$ supergravity. Let us note that in the truncation 
$\chi_i=0$ and upon normalization $2\varphi_0=-\varphi_1-\varphi_2$, the
remaining moduli $\{\varphi_i\}$ translate according to
\bea
x\equiv e^{-\varphi_1}\;,\quad y\equiv e^{-\varphi_2}
\;,
\eea
into the notation of \cite{DallAgata:2012tne} whose mass spectrum we reproduce.
Finally, as discussed in \cite{Catino:2013ppa}, when the moduli approach the boundary of the moduli space, e.g. $\varphi_{1,2} \to \pm \infty$, we obtain a different ${\cal N}=8$ gauged SUGRA. In particular, these limits can be understood as contractions of the gauge group to ${\rm SO}(2) \times {\rm SO}(2) \ltimes T^{26}$.

\section{Conclusions}

In this paper we used exceptional field theory to find the $D=10$ uplift of two inequivalent four-dimensional ${\cal N}=8$ gauged SUGRAs with the same dyonic gauge group ${\rm SO}(4)\times{\rm SO}(2,2) \ltimes T^{16}$, and which admit a non-supersymmetric Minkowski vacuum. We showed that the inequivalent four-dimensional theories come from truncating IIA or IIB around the same ${\rm Mink}_4 \times S^3 \times H^3$ background, with the IIA / IIB gaugings naturally arising in the ${\rm SU}^*(8)$ and ${\rm SL}(8)$ frames, respectively. The two consistent truncations are related by an outer automorphism of ${\rm SL}(4)$ which can be taken to act on the $S^3$, or $H^3$, using the techniques outlined in \cite{Malek:2015hma}.

The common ${\cal N}=4$ sector of these theories falls within the class considered in \cite{Baguet:2015iou}. By studying the ${\cal N}=4$ scalar potential we identified the four moduli which lie in the common NS-NS sector and parameterise the coset space $\left({\rm SL}(2)/{\rm SO}(2)\right)^2$. Using \cite{Baguet:2015iou} we uplifted these moduli to obtain a four-parameter family of Minkowski vacua in 10 dimensions which preserve a ${\rm U}(1)^4$ subgroup of the ${\rm SO}(4) \times {\rm SO}(2,2)$ isometries of the round $S^3 \times H^3$ background. Taking these scalar fields to the boundary of the moduli space results in new ${\cal N}=8$ gauged SUGRAs with gauge group ${\rm SO}(2) \times {\rm SO}(2) \ltimes {\rm T}^{26}$.

Gauged SUGRAs with dyonic gaugings are particularly interesting because of their rich vacuum structure, but have only recently been uplifted to 10-/11-dimensional SUGRA \cite{Guarino:2015jca,Guarino:2015qaa,Guarino:2015vca,Inverso:2016eet}. For example, half-maximal AdS vacua of type II and 11-dimensional SUGRA must have non-zero de Roo-Wagemans angles \cite{Louis:2014gxa,Malek:2017njj}. We hope that the techniques developed here will be useful in those applications.

Another interesting question raised by this work is whether our IIA uplift can be obtained directly in the ${\rm SU}^*(8)$ frame, as opposed to the commonly used ${\rm SL}(8)$ frame. This might lead to a generalisation of the IIA uplift, just as the IIB twist matrix is a particular example of a family of truncations obtained in \cite{Inverso:2016eet}. We leave these and other open questions for further work.

\subsection*{Acknowledgements}
We wish to thank Gianluca Inverso for very helpful discussions. We would also like to thank the Banff International Research Station for hospitality during the workshop ``String and M-theory geometries: Double Field Theory, Exceptional Field Theory and their Applications'' (17w5018), where part of this work was completed. EM would further like to thank ENS Lyon for hospitality. The work of EM is supported by the ERC Advanced Grant ``Strings and Gravity" (Grant No. 320045).

\providecommand{\href}[2]{#2}\begingroup\raggedright\endgroup

\end{document}